\documentclass[11pt,preprint,showpacs]{revtex4}

\usepackage{graphicx}
\usepackage{dcolumn}
\usepackage{bm}
\def\e{\kern+.6ex\lower.42ex\hbox{$\scriptstyle \iota$}\kern-1.20ex e}
\begin{document}

\title{Calculations of three-nucleon reactions with N$^3$LO chiral forces: 
achievements and challenges}

\author{H.~Wita{\l}a}
\affiliation{M. Smoluchowski Institute of Physics, Jagiellonian
University,  PL-30059 Krak\'ow, Poland}

\author{J.~Golak}
\affiliation{M. Smoluchowski Institute of Physics, Jagiellonian
University,  PL-30059 Krak\'ow, Poland}

\author{R.~Skibi\'nski}
\affiliation{M. Smoluchowski Institute of Physics, Jagiellonian
University,  PL-30059 Krak\'ow, Poland}

\author{K.~Topolnicki}
\affiliation{M. Smoluchowski Institute of Physics, Jagiellonian
University,  PL-30059 Krak\'ow, Poland}

\date{\today}

\begin{abstract}
 We discuss the application of the chiral N$^3$LO forces to
  three-nucleon reactions and point to the challenges which will have
  to be addressed.  
  Present approaches to solve three-nucleon Faddeev
  equations are based on a partial-wave decomposition. A rapid increase 
 of the number of terms
  contributing to the chiral three-nucleon force 
  when increasing the order of the chiral expansion from N$^2$LO to N$^3$LO 
 forced us to develop a 
  fast and effective method of automatized partial wave
  decomposition. At low energies of the 
  incoming nucleon below $\approx 20$~MeV, where only a limited
  number of partial waves is required,   
  this method allowed us to perform
  calculations of reactions in the three-nucleon continuum using
  N$^3$LO two- and three-nucleon forces.  
  It turns out that inclusion of consistent chiral interactions,
    with relativistic  $1/m$ corrections and short-range
  2$\pi$-contact term omitted  in the N$^3$LO three-nucleon force,    
 does not explain the long standing low energy $A_y$-puzzle. We discuss 
problems arising when chiral forces are applied at higher energies, where large 
 three-nucleon force effects are expected. 
 It seems plausible that at higher energies, due to a 
  rapid increase of a number of partial waves required to reach convergent
  results,  a  three-dimensional formulation of the Faddeev equations  which 
 avoids partial-wave decomposition is desirable.  
  
\end{abstract}


\maketitle
\setcounter{page}{1}

\section{Introduction}
\label{intro}
With the advent of nuclear forces derived in the framework of chiral
effective field theory a unique possibility has been offered 
to study few-nucleon
systems and their reactions with consistent two- and 
 many-nucleon interactions. A special place among
few-body systems is reserved for the three-nucleon (3N) system, for which
mathematically sound theoretical formulation in the form of the Faddeev
equations exists, both for bound and scattering states. In last
decades numerical algorithms have been developed which allow one to
 solve numerically  Faddeev equations
for any dynamical input, containing not only two- but also 
three-nucleon forces (3NF's) \cite{glo96,hub97}. 

Using these algorithms and standard, (semi)phenomenological
nucleon-nucleon interactions supplemented by model three-nucleon
forces, many investigations of the 3N continuum have been done in the past.  
High precision nucleon-nucleon potentials such as AV18~\cite{AV18},
CD~Bonn~\cite{CDBOnucleon-nucleon}, Nijm I and II~\cite{NIJMI},
which provide a very good description of the nucleon-nucleon data set up to
about 350 MeV, have been used.  
 They have been also combined with model 3N forces such as $2\pi$-exchange
 Tucson-Melbourne (TM99) 3NF \cite{TM99} or Urbana IX model \cite{uIX}.

When realistic NN  forces are used to predict binding energies
of three-nucleon systems they underestimate the experimental bindings
of $^3$H and $^3$He by about 0.5-1
MeV~\cite{Friar1993,Nogga1997}. This missing binding energy can be
restored by introducing a three-nucleon force into the nuclear
Hamiltonian~\cite{Nogga1997}. 
Also the study of elastic nucleon-deuteron (Nd) scattering and nucleon
induced deuteron breakup revealed a number of cases where the
nonrelativistic description using only pairwise forces is insufficient
to explain the data.  The best studied case at low energies is the
vector analyzing power in elastic nucleon-deuteron scattering for
which a large discrepancy exists in the region of its maximum around
c.m. angles $\theta_{c.m.} \approx 125^o$ and for incoming nucleon
energies below $\approx 20$~MeV \cite{glo96,wit01}.  
 For the elastic scattering angular
distribution at low energies negligible effects of 3NF's have been
found and theory based on realistic NN forces agrees well 
with the data \cite{glo96,wit01}.

That picture changes with increasing energy of the three-nucleon
system. Generally, the studied discrepancies between experiment and 
theory using only nucleon-nucleon (NN) potentials  become
larger and adding a
three-nucleon force to the pairwise interactions leads in some cases
to a better description of the data.  The elastic Nd 
angular distribution in the region of its minimum and at backward
angles is the best known example~\cite{wit98,sek02}.  The clear
discrepancy in these angular regions at energies up to
$\approx100$~MeV nucleon lab energy between a theory using only
 NN potentials and the cross section data can be removed
by adding a standard models of  three-nucleon forces to the nuclear
Hamiltonian. Such a 3NF must be adjusted with each
 NN potential separately to yield the experimental binding of
$^3$H and $^3$He~\cite{wit98,wit01,sek02}.  At energies higher than
$\approx 100$~MeV current three-nucleon forces only partially improve
the description of cross section data and the remaining discrepancies,
which increase with energy, indicate the possibility of relativistic
effects.  The need for a relativistic description of three-nucleon
scattering was also raised when precise measurements of the total
cross section for neutron-deuteron scattering~\cite{abf98} were
analyzed within the framework of nonrelativistic Faddeev
calculations~\cite{wit99}.  Nucleon-nucleon forces alone were
insufficient to describe the data above $\approx 100$~MeV.  The
effects due to relativistic kinematics considered in \cite{wit99} 
at higher energies were comparable in magnitude 
 to the effects due to three-nucleon
forces.  These results showed the importance of a study taking
relativistic effects in the three nucleon continuum into account.

In \cite{witrel1,witrel2} the first results on relativistic effects in
the three-nucleon continuum have been presented.  The dynamics was
defined by a three-nucleon total momentum zero frame Hamiltonian or mass
operator including only pairwise interactions.  The mass operator was
used to calculate three-nucleon scattering observables.  The input to
that approach is a ``Lorentz boosted'' nucleon-nucleon potential,
which generates the nucleon-nucleon $t$-matrix in a moving frame by
solving a standard relativistic Lippmann-Schwinger equation.  To get the
nucleon-nucleon potential in an arbitrary moving frame one needs the
interaction in the two-nucleon total momentum zero frame, which
appears in the relativistic nucleon-nucleon Schr\"odinger or
Lippmann-Schwinger equation.  The relativistic Schr\"odinger equation 
 in the two-nucleon total momentum zero frame 
 differs from the nonrelativistic Schr\"odinger equation just by the
relativistic form for the kinetic energy.  Current realistic
nucleon-nucleon potentials are defined and fit by comparing the
solution of the nonrelativistic Schr\"odinger equation to experimental
data.  Up to now nucleon-nucleon potentials refitted with the same
accuracy in the framework of the relativistic nucleon-nucleon
Schr\"odinger equation do not exist. Such refitting can be, however,
avoided by solving a quadratic integral equation whose solution is a
relativistic potential which is phase-equivalent to a given input
high-precision nonrelativistic nucleon-nucleon potential
\cite{kamada1}.

In our  studies with only nucleon-nucleon interactions we found
that when the non-relativistic form of the kinetic energy is replaced
by the relativistic one and a proper treatment of the relativistic dynamics is
included, the elastic scattering cross section is only slightly
 increased  by relativity at backward angles and higher energies 
 while spin observables are practically unchanged. 
 It is exactly the region of angles and energies where
the effects of three-nucleon forces are also significant
\cite{wit01}. 
 These observations prompted us to extend our 
 three-nucleon continuum
relativistic Faddeev calculations and include also three-nucleon
forces \cite{reland3nf}. 
 Again, only small relativistic effects at higher energies were found 
for the cross section  when 3NF had been included  and spin observables 
 remained practically unchanged. It supported conclusions from relativistic 
 calculations performed with 2N forces alone, that  for higher energies 
 discrepancies must reflect action of 3NF's.

The main drawback of all those studies was inconsistency between
applied NN interactions and 3N forces. In \cite{epel2002} for the 
 first time that
inconsistency was removed and low energy 3N continuum investigated
with chiral next-to-next-to-leading order (N$^2$LO) 
 NN and 3N forces. The NN interaction in that
order, however, does not describe the NN experimental  phase-shifts in
sufficiently wide energy range to allow application of those forces 
at higher energies. 

In \cite{mach_nn_n3lo} and \cite{epel_nn_n3lo} 
precise two-nucleon potentials have been
developed at next-to-next-to-next-to-leading order (N$^3$LO) of the chiral
expansion. They reproduce experimental phase-shifts 
\cite{nijm_phase1,nijm_phase2} in a wide
energy range and practically with the same high precision as realistic
(semi)phenomenological NN potentials. The necessary work to derive the
 consistent chiral 3NF's at N$^3$LO has been done in \cite{3nf_n3lo_long} 
and \cite{3nf_n3lo_short}. In that
order five different topologies contribute to the 3NF. Three of them
 are of   long-range character \cite{3nf_n3lo_long} 
 and are given by two-pion (2$\pi$)
exchange graphs, by two-pion-one-pion (2$\pi$-1$\pi$) exchange graphs,
and by the so-called ring diagrams. They are
supplemented by the short-range two-pion-exchange-contact (2$\pi$-contact)
 term and by the leading relativistic corrections to the
three-nucleon force \cite{3nf_n3lo_short}. 

Results of 
refs. \cite{mach_nn_n3lo,epel_nn_n3lo,3nf_n3lo_long,3nf_n3lo_short} 
  enable one now to perform for the first time consistent
calculations of three-nucleon reactions at N$^3$LO order of chiral  
expansion.   The 3NF at this
order does not involve any new unknown low-energy constants (LECs)
and depends only on two parameters, $c_D$ and $c_E$, that parametrize the
leading one-pion-contact term and the three-nucleon contact term
appearing at N$^2$LO.  Their values need to be fixed at given order 
 from a fit to
few-nucleon data. Among the few possible observables that have been
used in this connection are the triton binding energy and the
 neutron-deuteron doublet scattering length $^2a_{nd}$ \cite{epel2002}, the
$^4$He binding energy, the properties of light nuclei, or the triton
$\beta$ decay rate. 

Application of N$^3$LO 3NF in few-body calculations is challenging due
to its very rich and complicated operator structure. The large number of
terms in the 3NF at N$^3$LO \cite{3nf_n3lo_long,3nf_n3lo_short}  
 requires an effective method of
performing partial-wave decomposition. Recently such a method, which
comes under the name of automatized partial-wave decomposition (aPWD) was
proposed by us in \cite{apwd,apwd_a}. In that approach 
 the matrix elements in the 3N momentum-space
partial wave basis  for different terms
contributing to N$^3$LO 3NF are obtained in two consecutive steps. In
the first the spin-momentum and isospin parts of three-nucleon
interactions are calculated using a software for symbolic
calculations. The resulting momentum-dependent functions are then
integrated numerically in five dimensions over angular variables. The
major advantage of this method is its generality since it can be applied
to any momentum-spin-isospin operator. 

In the present paper we would like to present first results obtained 
with N$^3$LO chiral forces and discuss problems which  
should be resolved in the future. 
 In section \ref{low} results for low energy elastic 
 neutron-deuteron (nd) scattering will be shown, followed by low 
energy breakup. 
 In section \ref{form} we present results  and 
 discuss problems  encountered at higher energies. 
 To resolve some of those problems and to solve 3N Faddeev equations 
with NN and 3N chiral forces included  it seems that at higher 
energies three-dimensional approach which avoids partial-wave decomposition 
 is more adequate  and in section \ref{3dim} we describe the present state of 
such an approach.
 We summarize in section \ref{summary}.

\section{Application of N$^3$LO chiral forces at low energy}
\label{low}

\subsection{Elastic scattering}
\label{elastic}

The nuclear Hamiltonian at N$^3$LO order of chiral expansion  is fixed by 
values of LEC's $c_D$ and  $c_E$. 
To determine them we follow ref. \cite{epel2002}
and use the experimental triton binding energy $E^{^3H}$ and the 
 nd doublet scattering length $^2a_{nd}$ as two observables from which
$c_D$ and $c_E$ can be obtained. The procedure can be divided into two
steps. First, the dependence of $E^{^3H}$ on $c_E$ for a given
value of $c_D$ is determined. The requirement to reproduce the
experimental value of the triton binding energy yields a set of
combinations $c_D$ and $c_E$. This set is then used in the
calculations of $^2a_{nd}$, which allows us to find which pair of
$c_D$ and $c_E$ describes both observables simultaneously. 

We compute the $^3$H wave function using the method described in
\cite{Nogga1997}, where the full triton wave function $\Psi = (1+P) \psi$ is
given by its Faddeev component $\psi$, which fulfills the Faddeev
equation
\begin{eqnarray}
\psi = G_0tP\psi + (1+G_0t)G_0V^{(1)}(1+P)\psi .
\label{eq1}
\end{eqnarray}
Here $G_0$ is the free 3N propagator, P is the permutation operator, 
t is the two-body t-matrix generated from a given NN potential
through the Lippmann-Schwinger equation, and $V^{(1)}$ is a part of a 3NF
symmetric under the exchange of nucleons 2 and 3. 

The doublet scattering length $^2a_{nd}$ is calculated using 
($c_D$,$c_E$) pairs, which reproduce the correct value of
$E^{^3H}$. To this end we solve the Faddeev equation for the auxiliary
state $ T \phi $ at zero incoming energy \cite{zeroenergy}
\begin{eqnarray}
T \phi &=& tP\phi + (1+tG_0)V^{(1)}(1+P)\phi + tPG_0T \phi +
(1+tG_0)V^{(1)}(1+P)G_0T \phi ,
\label{eq2}
\end{eqnarray}
where the initial channel state $\phi$ occurring in the driving terms
is composed of the deuteron and a plane-wave state of the projectile
nucleon. The amplitude for the elastic scattering leading to a corresponding 
final state $\phi ' $ is then given by
\begin{eqnarray}
\phi' U \phi &=&  \phi' PG_0^{-1} \phi + 
 \phi' PT \phi  + \phi' V^{(1)}(1+P)\phi 
+ \phi' V^{(1)}(1+P)G_0T\phi ,
\label{eq3}
\end{eqnarray}
and for the breakup reaction reads
\begin{eqnarray}
\phi_0' U_0 \phi  &=& \phi_0'  (1 + P)T \phi ,
\label{eq3_br}
\end{eqnarray}
where $\phi_0'$ is the free three-body channel state. 
We refer to  \cite{glo96} and \cite{hub97} for a general overview of
3N scattering and for more details on the practical implementation of
the Faddeev equations.

\begin{figure}
\includegraphics[scale=0.7]{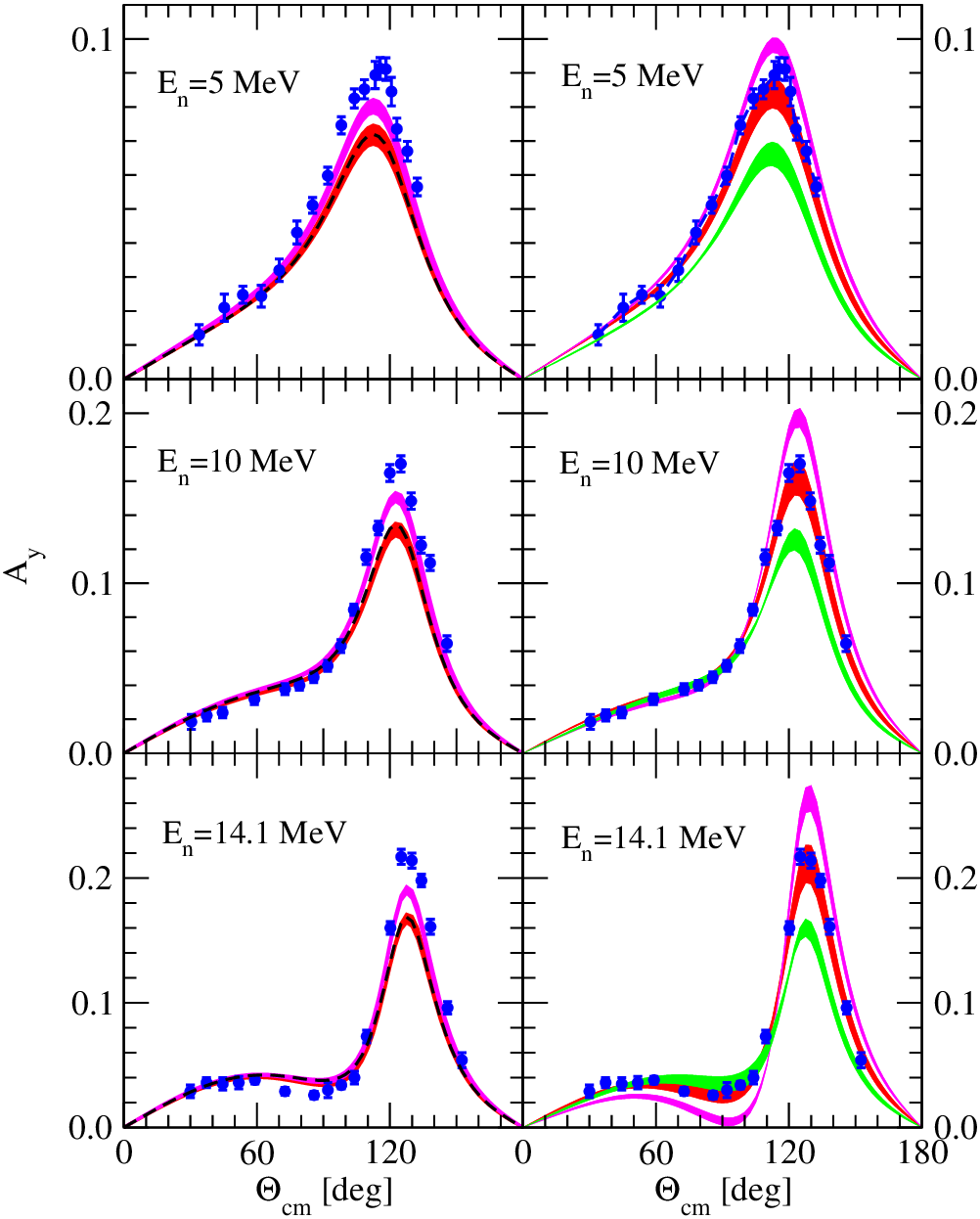}
\caption{
(color online)
The neutron
analyzing power $A_y$ in elastic nd scattering. In the left column
the dark shaded (red) and light shaded
(magenta) bands show predictions of realistic
NN potentials (AV18, CD~Bonn, Nijm1 and Nijm2) alone or combined with
the TM99 3NF, respectively. The dashed (black) line shows prediction of 
AV18 + Urbana IX combination. In the right column the magenta (upper), red
(middle) and green (low)
bands show predictions of the Bochum NLO, N$^2$LO, and N$^3$LO chiral NN
potentials, respectively, for the five cut-off parameters from 
 Table~\ref{table1}.
  The nd data (full blue circles) at $5$~MeV are from \cite{tunl1}, 
at $10$~MeV from \cite{tunl2}, and at $14.1$~MeV from \cite{tunl3}. 
}
\label{fig1}
\end{figure}

In this first preliminary study we restrict the application of N$^3$LO 3NF to 
 nd reactions at low energies, with incoming neutron lab. energy 
below $\approx 20$~MeV. 
To get converged results at such energies it is sufficient to include 
2N force components with a total two-nucleon angular momenta $j \le 3$ 
 in 3N partial-wave states with the total 3N system angular momentum 
$J \le 25/2$. Including 3NF it is sufficient  
 to incorporate its matrix elements with $j \le 3$ and $J \le 5/2$. 
At those energies the most interesting observable is the analyzing
  power $A_y$ for nd elastic scattering with polarized neutrons. 
 Theoretical predictions of standard, high precision NN  
potentials cannot explain the data for $A_y$. The data are underestimated 
by $\approx 30 \%$ in the region of $A_y$ maximum which occurs 
in region of c.m. angles $\Theta_{cm} \approx 120 ^o$. Combining standard NN 
potentials  
with commonly used models of a 3NF such as TM99 or Urbana IX models
 removes approximately only half of the discrepancy with respect to the 
data (see Fig.~\ref{fig1}). 

\begin{figure}
\includegraphics[scale=0.7]{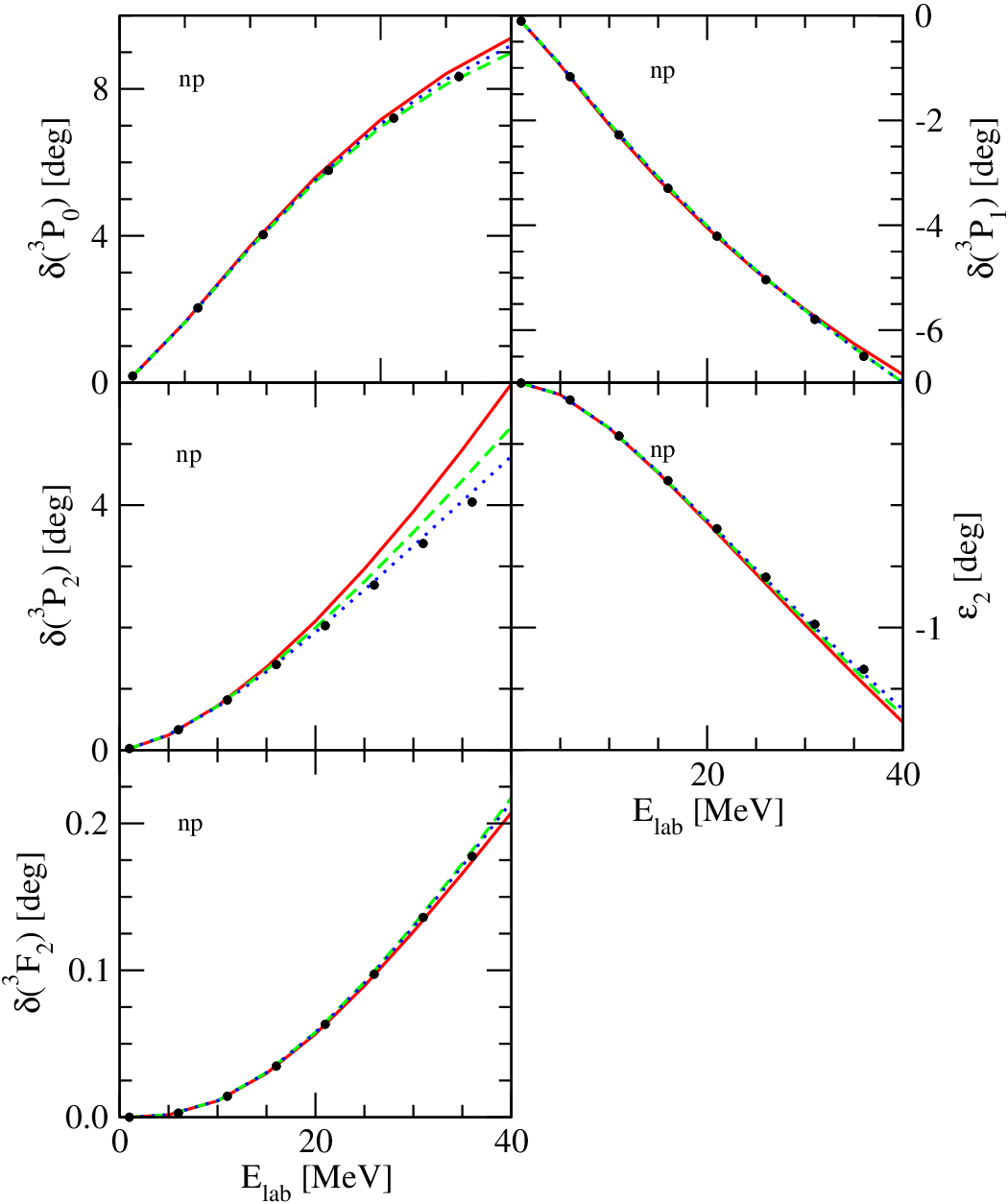}
\caption{
(color online)
The neutron-proton $^3P_j$ phase-shifts as 
 a function of lab. energy $E_{lab}$. 
The solid (red), dashed (green), and dotted (blue) lines show predictions of 
the Bochum NLO, N$^2$LO, and N$^3$LO NN potentials with a second cut-off 
parameter of Table~\ref{table1}, 
respectively. The solid (black) circles are experimental Nijmegen phase-shifts 
\cite{nijm_phase1,nijm_phase2}.
}
\label{fig2}
\end{figure}

When instead of standard forces chiral NN interactions are used, the 
predictions    for $A_y$ vary with the order of chiral expansion. 
Theory based on NLO interactions clearly overestimates the $A_y$ data 
while N$^2$LO forces give quite a good 
 description of them (see Fig.~\ref{fig1})
 leading thus to disappearance of $A_y$-puzzle. Only when N$^3$LO NN chiral 
forces are used, the picture resembles again the one for standard forces, with 
clear discrepancy between theory and data in the region of $A_y$ maximum 
(see Fig.~\ref{fig1} where bands of predictions for five versions of 
the Bochum NLO, N$^2$LO and N$^3$LO potentials with cut-off parameters from 
Table \ref{table1} are shown). 
Such behaviour can be traced back to a high sensitivity 
of $A_y$ to $^3P_j$ NN force components and to the fact, that only at N$^3$LO 
order of chiral expansion the experimental $^3P_j$ phases 
 \cite{nijm_phase1,nijm_phase2}, 
especially $^3P_2$-$^3F_2$, are properly 
 reproduced (see also Fig.~\ref{fig2}). 
 
\begin{figure}
\includegraphics[scale=0.7]{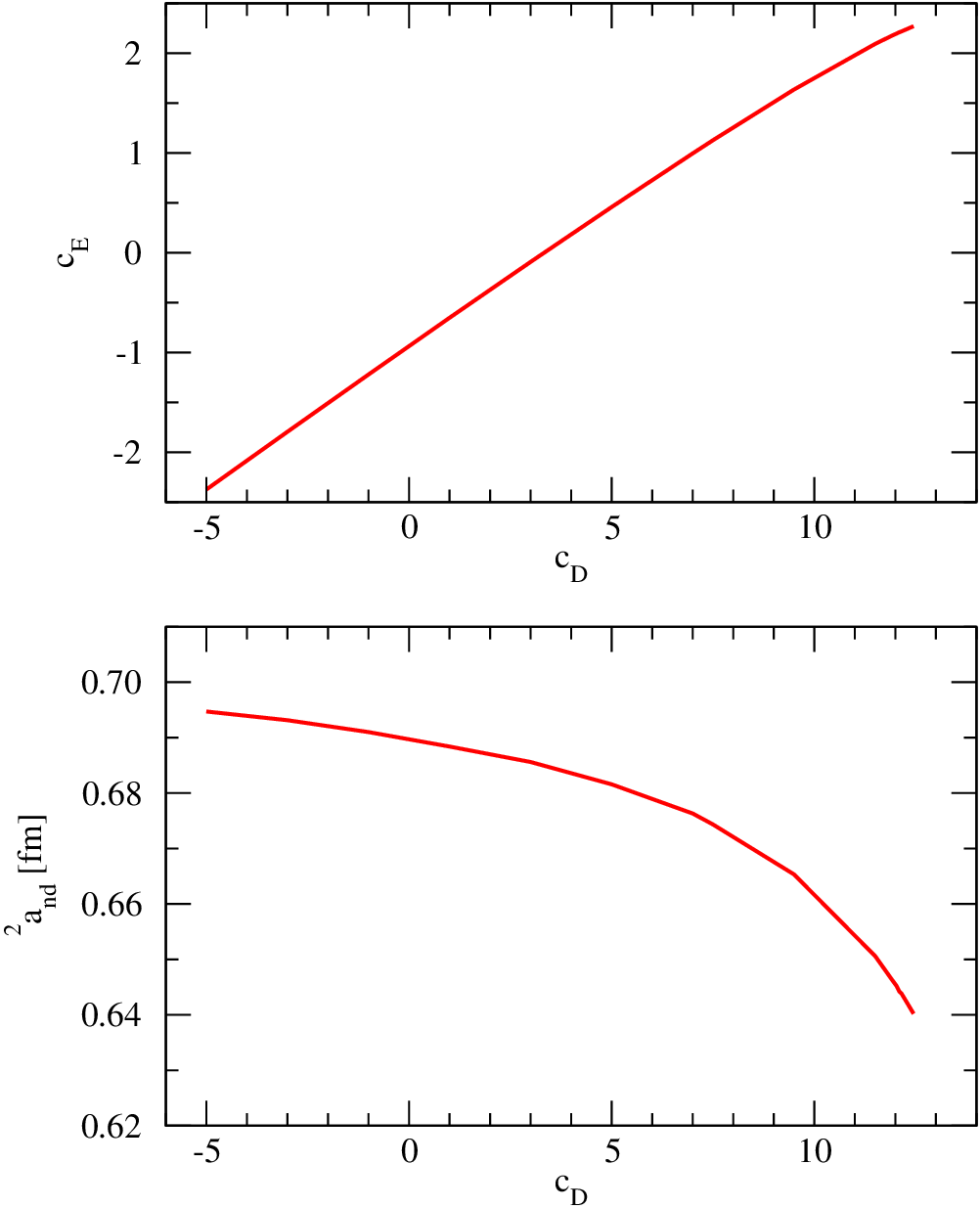}
\caption{
(color online)
Values of ($c_D$, $c_E$) pairs which reproduce $^3$H experimental binding
 energy (upper part) and dependence of the doublet nd scattering 
length $^2a_{nd}$ on $c_D$. The Bochum N$^3$LO chiral 2N- and 3N-forces 
were used in calculations with a second cut-off parameter 
 from Table~\ref{table1}. 
}
\label{fig3}
\end{figure}

\begin{figure}
\includegraphics[scale=0.7]{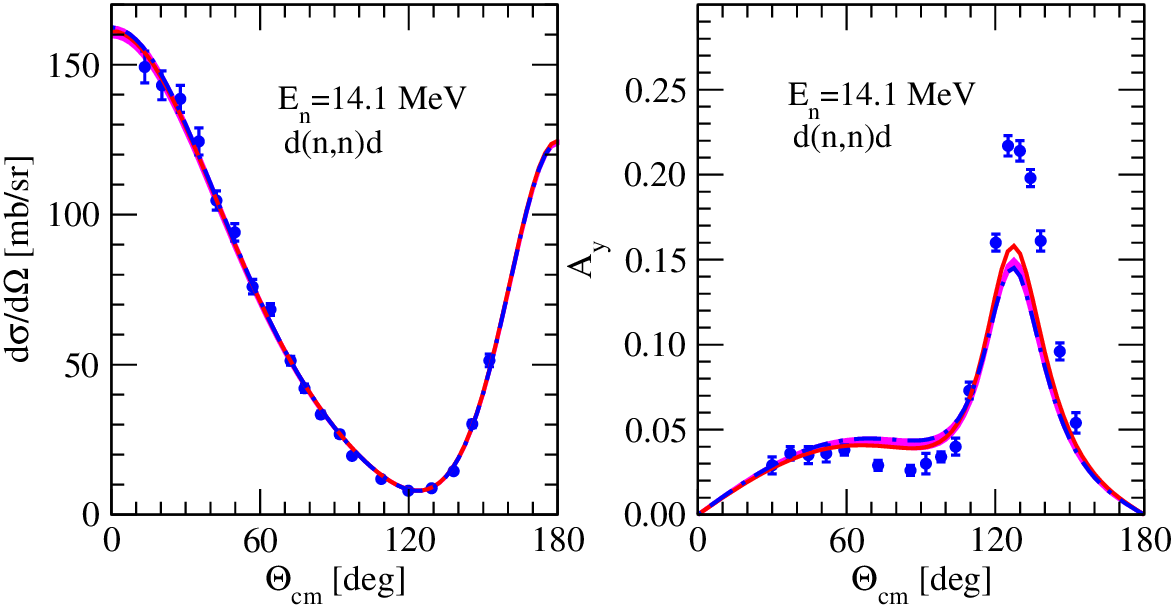}
\caption{
(color online)
The neutron cross section (left part) and 
analyzing power $A_y$ (right part) 
for elastic nd scattering at $E_n=14.1$~MeV. 
 The solid (red) line shows predictions of the Bochum N$^3$LO NN
potential (with the second cut-off parameter from Table~\ref{table1}) 
 alone and the dashed-dotted (blue) line results 
when that potential  is combined with the
 N$^3$LO 3NF without  1/m corrections and $2\pi$-exchange-contact
 term. The light shaded 
(magenta) bands show predictions for that combination of NN + 3NF forces with 
 values of $c_D$ and $c_E$ parameters from upper part of Fig.~\ref{fig3}. 
 The nd data (full blue circles) for the cross section are from \cite{berick} 
and for $A_y$  from \cite{tunl3}.
}
\label{fig4}
\end{figure}

The question arises if consistent chiral N$^3$LO 3NF's can provide 
explanation for the low energy $A_y$-puzzle. In this first investigation 
we included all long-range contributions to N$^3$LO 3NF with the exception
of 1/m corrections. Additionally, the $2\pi$-exchange-contact term
was omitted in the short-range part of 3NF.
 In Fig.~\ref{fig3} we show 
for the second cut-off parameter from Table~\ref{table1}
  a typical dependence of $c_E$ and $c_D$ 
parameters which reproduce the experimental triton binding energy. The 
requirement to reproduce in addition also $^2a_{nd}$ scattering length leads to
 $c_E$ and $c_D$  values shown in Table~\ref{table1} and $A_y$ values shown 
in Fig.~\ref{fig4} by dashed-dotted (blue) line. 
It turns out that adding that N$^3$LO 3NF does not improve 
the description of $A_y$; on the contrary, even lowers slightly the maximum 
of $A_y$ increasing thus a discrepancy between the theory and the data.

In order to check how restrictive is the requirement to reproduce 
in addition to $^3$H binding energy also the 
experimental value of $^2a_{nd}$ we show in Fig.~\ref{fig4} also a band 
of predictions for ($c_E$, $c_D$) pairs from Fig.~\ref{fig3}. 
 Since that band is  narrow it implies
that even when allowing for more freedom, that is 
without reproducing of $^2a_{nd}$,
the $A_y$-puzzle cannot be explained by that N$^3$LO 3NF. 

\begin{figure}
\includegraphics[scale=0.7]{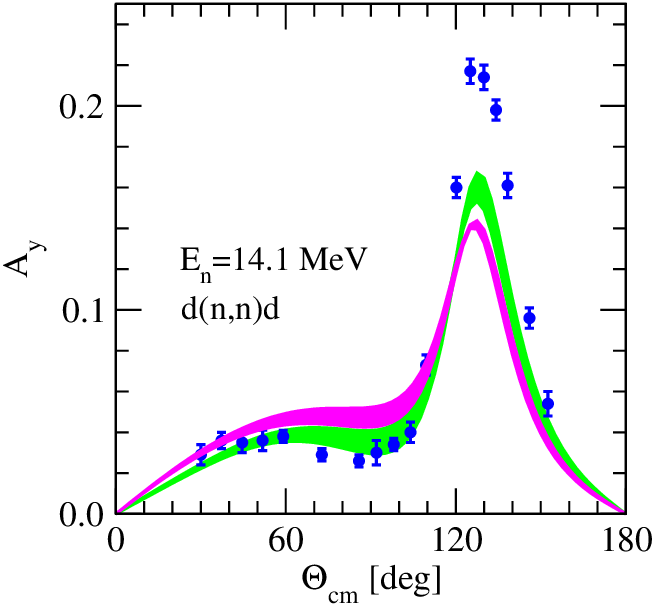}
\caption{
(color online) 
The neutron
analyzing power $A_y$ elastic nd scattering at $E_n=14.1$~MeV. 
The light shaded (green)  and dark shaded (magenta) bands  contain 
predictions of the Bochum N$^3$LO NN
potentials with five cut-off parameters of Table~\ref{table1} alone and 
when they are  combined with
 N$^3$LO 3NF without  $2\pi$-exchange-contact term, respectively.
 The nd data (full circles) are from \cite{tunl3}.
}
\label{fig5}
\end{figure}

In Fig.~\ref{fig5} we show   bands of predictions for $A_y$ using 
 N$^3$LO chiral NN potentials for five values of 
cut-off's and combining them with N$^3$LO 3NF with  ($c_E$, $c_D$) pairs 
 from Table~\ref{table1}. It is clear 
 that N$^3$LO 3NF without 1/m corrections and $2\pi$-contact term 
 is not able to explain the $A_y$-puzzle. 

The results for the cross section at incoming neutron energy $14.1$~MeV  
 shown in the left part of  Fig.~\ref{fig4} 
 exemplify negligible 3NF effects for the elastic scattering cross section   
at low energies.

\subsection{Breakup}
\label{breakup}

\begin{figure}
\includegraphics[scale=0.7]{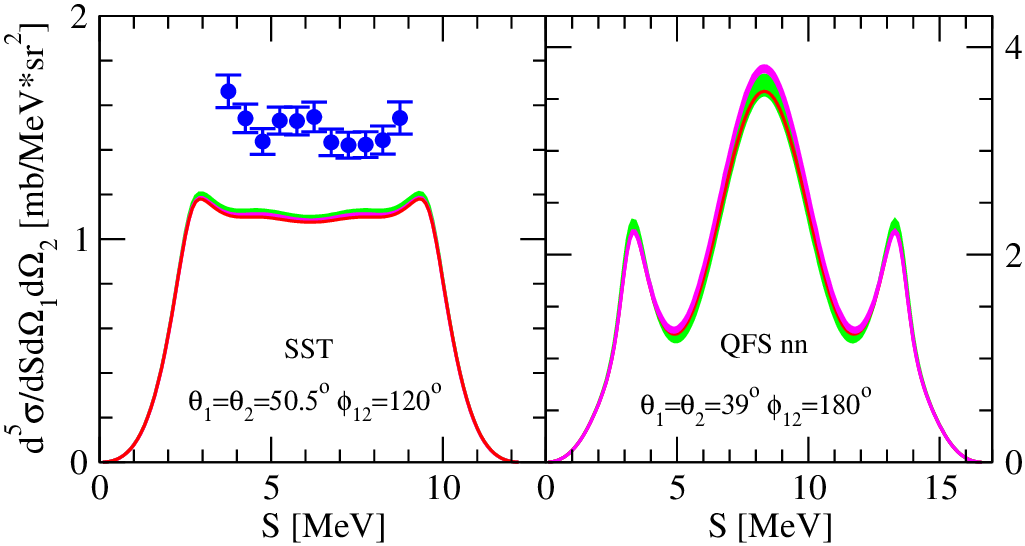}
\caption{
(color online) 
The cross section 
  $d^5\sigma/d\Omega_1d\Omega_2dS$ 
for the d(n,nn)p breakup reaction
as a function of the 
  arc-length S at $E_n^{lab}=13$~MeV 
 for the SST
  and QFS nn configurations.  
 The light shaded (green) and dark shaded
(magenta) bands show predictions of the Bochum  N$^3$LO NN
potentials alone and combined with the N$^3$LO 3NF (without short-range
$2\pi$-exchange-contact term) for five different cut-offs 
 of Table~\ref{table1},
 respectively.
 The solid (red) line is a prediction obtained with the  CD~Bonn potential.
 The full (blue) circles are nd data for the SST 
 configuration  from \cite{sst}.
}
\label{fig6}
\end{figure}

Cross sections for the symmetric-space-star (SST) and
quasi-free-scattering (QFS) configurations of the nd breakup
are very stable with respect to the underlying dynamics.
Different potentials, alone or combined
with standard 3N forces, provide practically the same SST and QFS
 cross sections \cite{din1}. Also the chiral
 N$^3$LO 3NF without relativistic  $1/m$ corrections and short-range
  2$\pi$-contact term is no exception and  cannot explain the discrepancy 
 between the theory 
and the data found for the SST configuration \cite{sst} (Fig.\ref{fig6}).
 At low energies  the cross sections in the SST and QFS configurations are
 dominated by the S-waves. For the SST configuration the largest
contribution to the cross section comes from the $^3S_1$ partial 
 wave, while for
neutron-neutron (nn) QFS
 the $^1S_0$ partial wave dominates.
Neglecting rescattering, the QFS configuration resembles free NN
scattering. For free, low-energy neutron-proton (np) scattering one expects
contributions from $^1S_0$ np and $^3S_1$ force components. For free nn
scattering only the $^1S_0$ nn channel is allowed. That implies that nn QFS
would be a powerful tool to study the nn interaction.
The measurements of np QFS cross sections  have shown good agreement
 of data with theory \cite{exqfs}, confirming thus good knowledge of
 the np force.
For nn QFS it was found that theory underestimates the data by
$\approx 20\%$ \cite{exqfs}. The large
stability of the QFS cross sections
 with respect to the underlying dynamics, means that the present day
 $^1S_0$ nn interaction is probably incorrect \cite{din1,din2,din3}.

\section{Higher energies}
\label{form}

\begin{figure}
\includegraphics[scale=0.7]{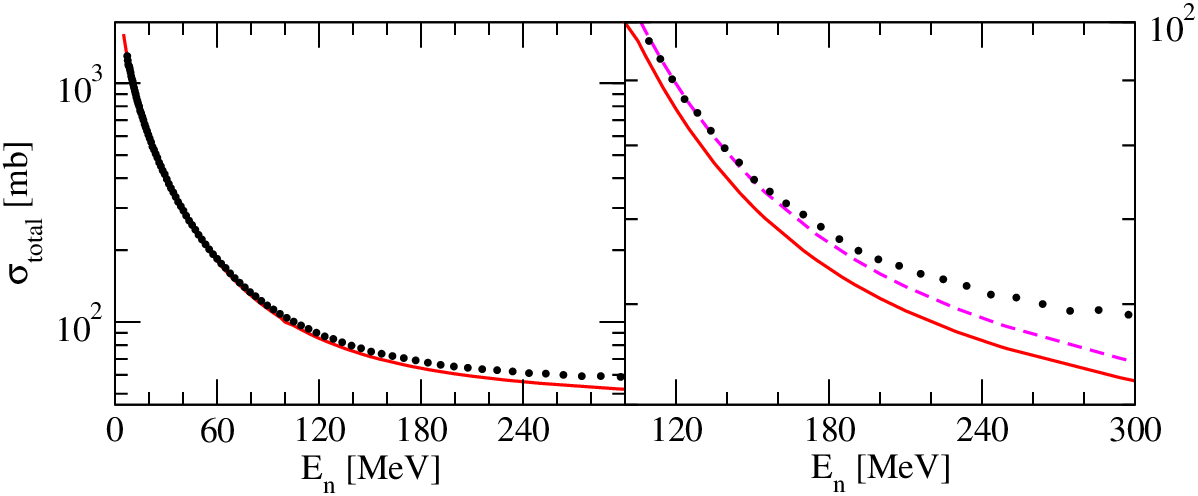}
\caption{
(color online) 
The total nd cross section as a function of the neutron lab. energy $E_n$. 
 The solid (red) and dashed (magenta) lines show predictions of the CD~Bonn 
 potential alone and combined with TM99 3NF, respectively.  
The black dots are nd data from  \cite{abf98}.
}
\label{fig7}
\end{figure}

\begin{figure}
\includegraphics[scale=0.7]{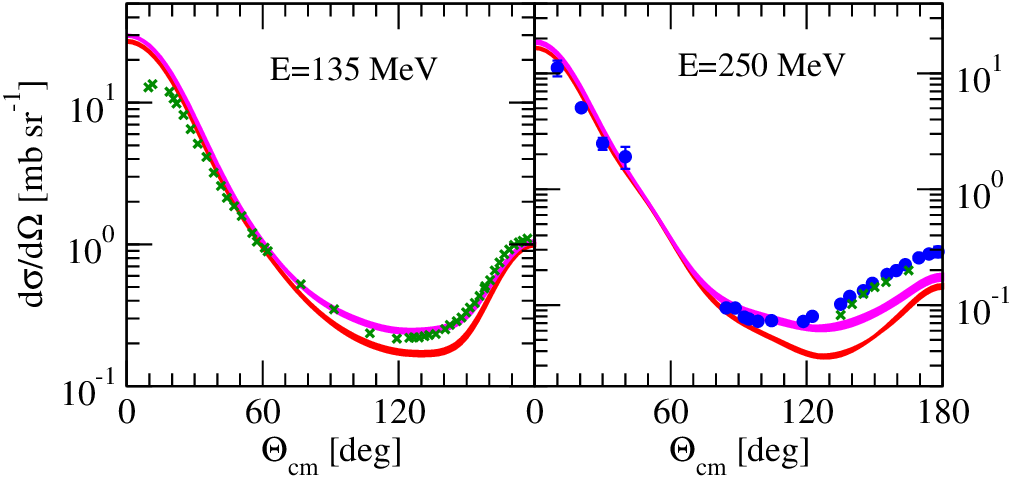}
\caption{
(color online) 
The  nd elastic scattering angular distributions at $135$~MeV and $250$~MeV 
of incoming neutron lab. energy. The dark shaded (red) and the light 
shaded (magenta) bands are predictions of 
standard NN potentials (AV18, CD~Bonn, Nijm I and II) alone and when 
they are combined with the TM99 3NF, respectively. 
The (green) x-es are pd data  at $135$~MeV from \cite{sek02} and at $250$~MeV  
from \cite{hat02}. Full (blue) circles are nd data from \cite{maedand}.
}
\label{fig8}
\end{figure}

Studies performed with standard NN potentials revealed that 
clear discrepancies between theory and data occur
for increasing 
incoming nucleon energy. They start to appear 
 around $E_N^{lab} \approx 60$~MeV 
for elastic scattering cross sections and spin 
observables and their magnitude 
grows with the energy of a 3N system \cite{wit01,abf98}. 
 Fig.\ref{fig7} and Fig.\ref{fig8} exemplify them for 
the total nd interaction cross section and elastic scattering angular 
distributions, respectively. Combining standard NN potentials with commonly 
used 3NF's such as TM99 or Urbana IX models, leads to 
a  good  description of data for cross sections up to $\approx 130$~MeV 
(see Fig.\ref{fig7} and Fig.\ref{fig8}), fails however completely at higher 
energies. Also for spin observables at higher energies a complex pattern of 
angular and energy discrepancies between theory based on standard nuclear 
forces and data was revealed. That pattern 
 cannot be explained even when standard 
models of 3NF's are included into 
calculations \cite{sek02,sek_eltransfer,hat02}. 3N continuum relativistic 
calculations have shown that relativistic effects, even 
when 3NF's are included, are negligible for the elastic scattering spin 
observables and raise the elastic cross section 
 at backward angles  \cite{reland3nf}
only slightly at higher energies. 
 It implies that large discrepancies between theory and data 
 as well as their complex pattern at higher energies are caused 
 by other terms contributing to 3NF, possibly  
 short  range 3NF components. 
 With the increasing energy of the 3N system they play a more and more
important 
role. Consistent N$^3$LO NN and 3N forces provide thus a unique possibility 
to check if such a scenario of increasing importance of different  3NF terms 
is able to provide a good description of observables in 3N reactions at 
higher energies.

The basic question is if the application of chiral forces to 3N continuum 
can be justified  and extended to energies as high as 200 MeV of incoming 
nucleon lab. energy ?
 At that energy  the value of the expansion parameter 
 $p/\Lambda \approx 0.45 $ therefore  application of chiral forces
 should be possible. 
 However, before taking this step, problems arising from the cut-off dependence
  of chiral N$^3$LO NN force for higher energy predictions  
  should be resolved.

\begin{figure}
\includegraphics[scale=0.7]{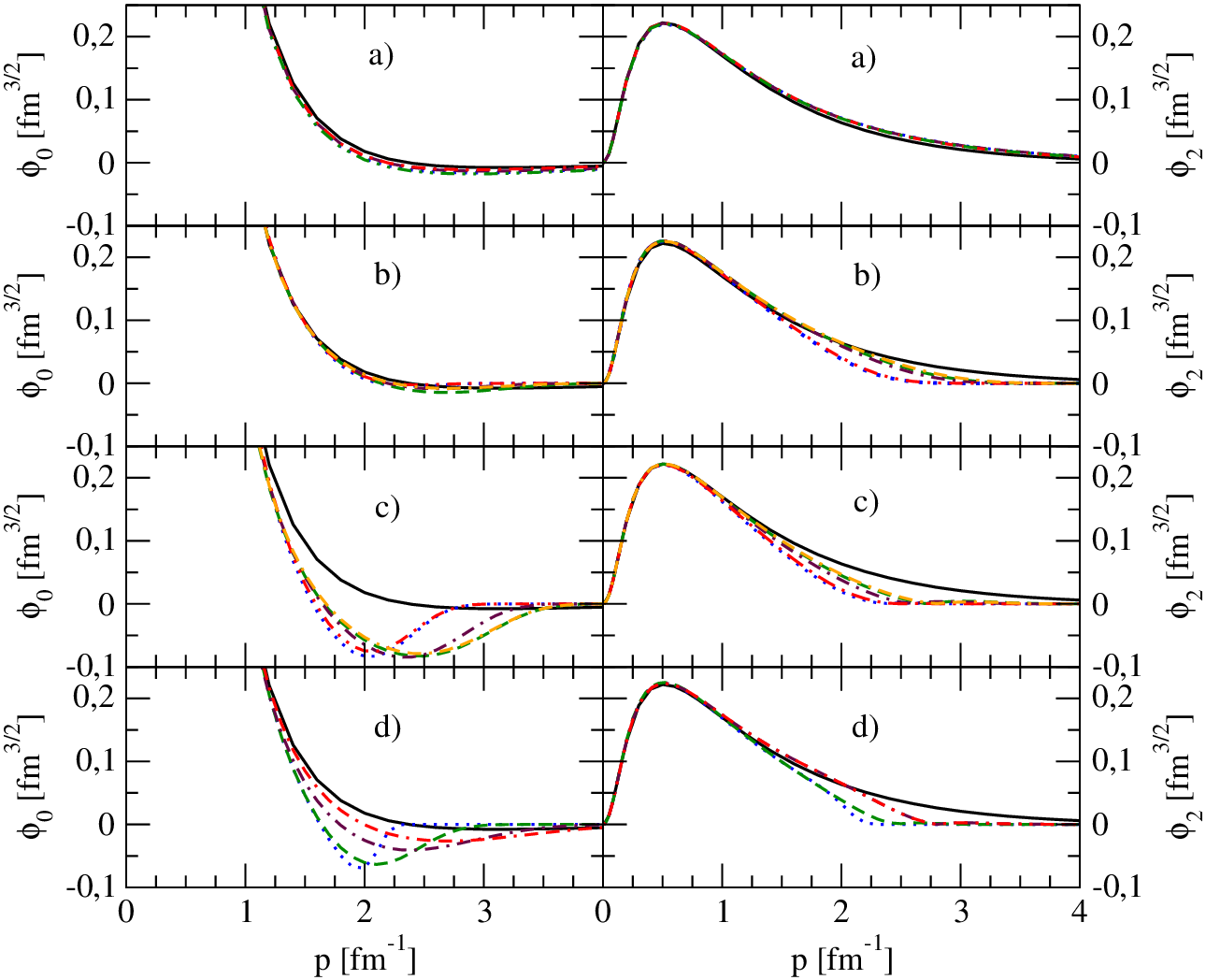}
\caption{
(color online) 
The momentum space deuteron wave function for different NN potentials. 
 The S- and D-components ($\phi_0$ and $\phi_2$, respectively) 
 are shown in the left and right columns, 
respectively.  
In a) wave functions for standard NN potentials are shown by different lines: 
AV18 - dotted (blue), CD~Bonn - solid (black), Nijm 93 - dashed (maroon), 
Nijm I - dashed-dotted (red), and Nijm II - dashed-double-dotted (green). 
In b) and c) wave functions for the  Bochum N$^2$LO 
 and N$^3$LO NN potentials \
with different cut-off parameters from Table \ref{table1} are shown, 
respectively: 
(450,500) - dotted (blue) line, 
(600,500) - dashed (green) line, 
(550,600) - dashed-dotted (maroon) line, 
(450,700) - dashed-double-dotted (red) line, 
(600,700) - double-dashed-dotted (orange) line. 
 The Idaho N$^3$LO wave functions for different cut-off parameters  
from Table \ref{table2} are shown in d): 
414 - dotted (blue) line, 
450 - dashed (green) line, 
500 - dashed-dotted (maroon) line, 
600 - dashed-double-dotted (red) line. 
For comparison in b), c) and d) also the CD~Bonn wave function 
is shown by solid (black) line.
}
\label{fig9}
\end{figure}

\begin{figure}
\includegraphics[scale=0.7]{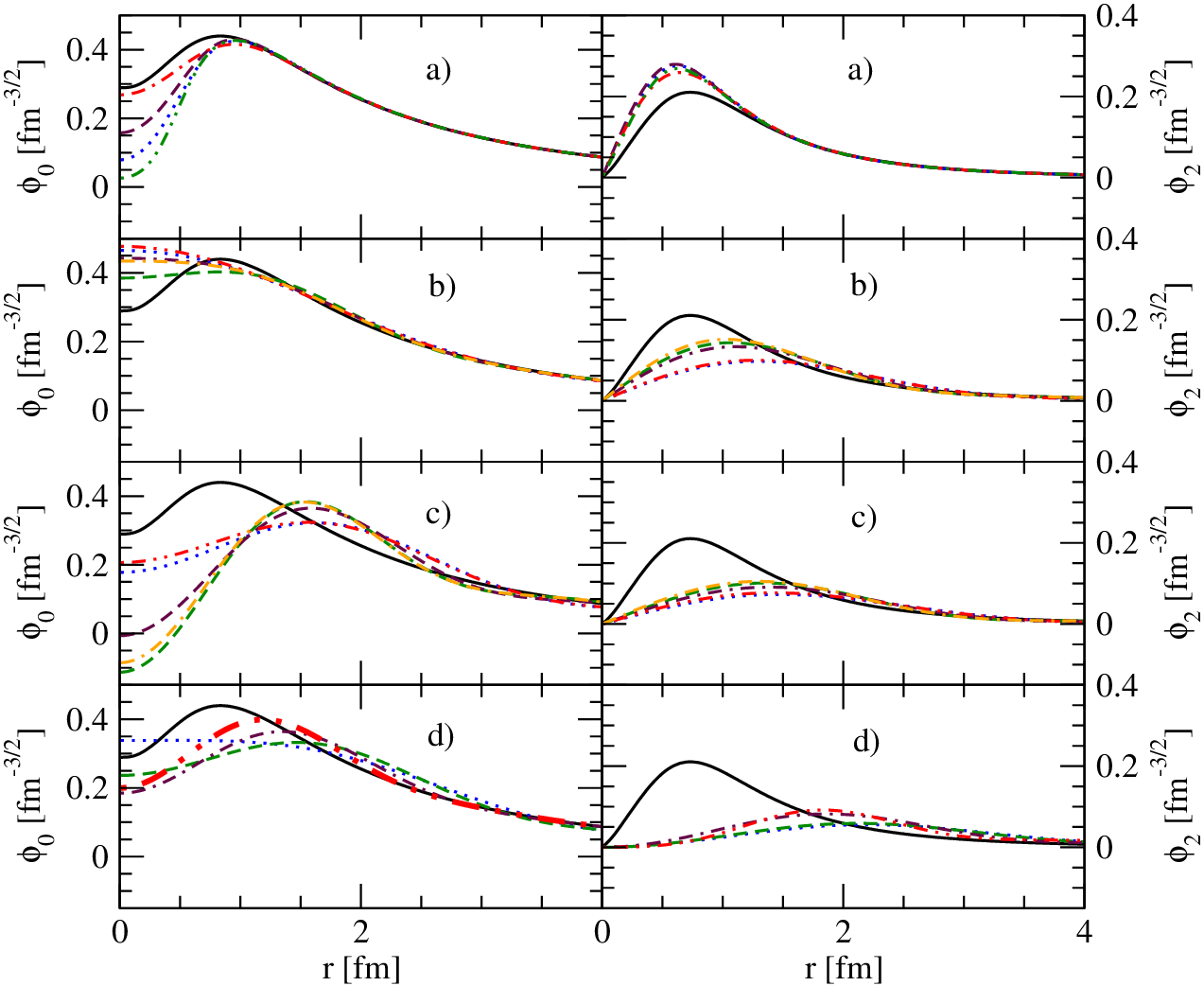}
\caption{
(color online) 
The coordinate space deuteron wave function for different NN potentials.
For explanation of lines see Fig.\ref{fig9}.
}
\label{fig10}
\end{figure}

A number of 
 N$^3$LO potentials with different cut-off parameter ranging 
from $414 -700$~MeV  (see Table \ref{table1} and \ref{table2}) were 
developed by the Bochum \cite{epel_nn_n3lo} and 
 Idaho \cite{mach_nn_n3lo} groups, 
 which equally 
well describe the experimental NN phase-shifts up to $\approx 200$~MeV 
with the same high precision as standard  NN potentials. 
However, changing the cut-off in such a wide range of values leads to 
deuteron wave functions which vary not only between themselves 
but also with respect to deuteron wave functions of 
 standard NN potentials at momenta 
 around $p \approx 2$~fm$^{-1}$ (see Fig.\ref{fig9}) or distances around    
$r \approx 1.5$~fm (see Fig.\ref{fig10}). Since nd elastic scattering 
transition amplitude Eq.~(\ref{eq3}) contains an 
 exchange term $ \phi'  PG_0^{-1} \phi $ 
determined by a deuteron wave function, it is necessary to check to what extent
the variations of a deuteron wave function with a cut-off parameter 
are reflected in elastic scattering observables. 
 With increasing energy of the incoming nucleon, the relative nucleon-deuteron 
momentum $q_0$ grows, taking value $q_0 = 2.07$~fm$^{-1}$ 
 at $E_N^{lab}=200$~MeV. Exactly at this value of momentum the largest 
variations of the deuteron wave function for different NN N$^3$LO
chiral forces occur (see Fig.\ref{fig9}). 

To study the influence of different cut-offs and corresponding deuteron 
wave functions on elastic scattering observables 
we solved the 3N Faddeev equation at three energies: $65$~MeV, 
$135$~MeV, and $200$~MeV, with 
the Bochum N$^3$LO (five cut-off values of Table \ref{table1})  and 
 N$^3$LO Idaho (four cut-off values of Table \ref{table2}) NN potentials. 
In order to check how results change with the order of the chiral expansion 
we also performed such calculations with 
 the Bochum N$^2$LO (five cut-off values 
of Table \ref{table1}) NN interactions. 

\begin{figure}
\includegraphics[scale=0.7]{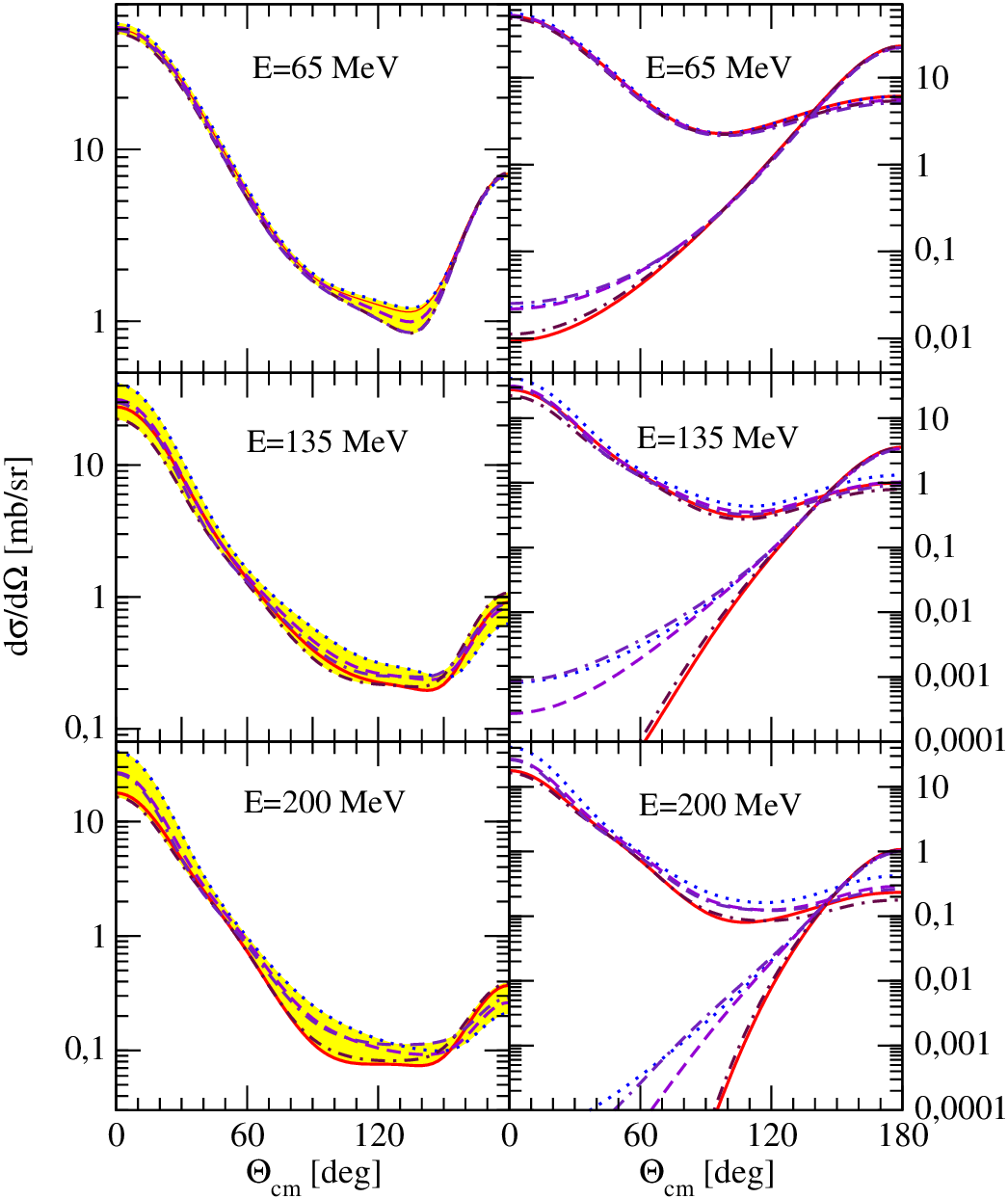}
\caption{
(color online) 
The  nd elastic scattering angular distributions at $65$~MeV, $135$~MeV 
and $200$~MeV 
of incoming neutron lab. energy calculated with 
 the Bochum N$^2$LO NN potentials 
 for different cut-off values of Table \ref{table1}. 
In the left part the nd elastic scattering cross sections are shown. In 
 the right part ``cross sections'' resulting only from the $PT$ term  
in the elastic scattering transition amplitude are shown (lines peaked at 
forward angles)  together with ``cross sections'' based on 
 exchange-term $PG_0^{-1}$ only (lines peaked at backward angles). 
Different lines correspond to different cut-off parameters 
from Table \ref{table1}: 
(450,500) - solid (red), 
(600,500) - dotted (blue), 
(550,600) - dashed (violet), 
(450,700) - dashed-dotted (maroon), 
(600,700) - double-dashed-dotted (indigo). 
In the left part light-shaded (yellow) bands show scatter of predictions 
for different cut-off values.  
}
\label{fig11}
\end{figure}

\begin{figure}
\includegraphics[scale=0.7]{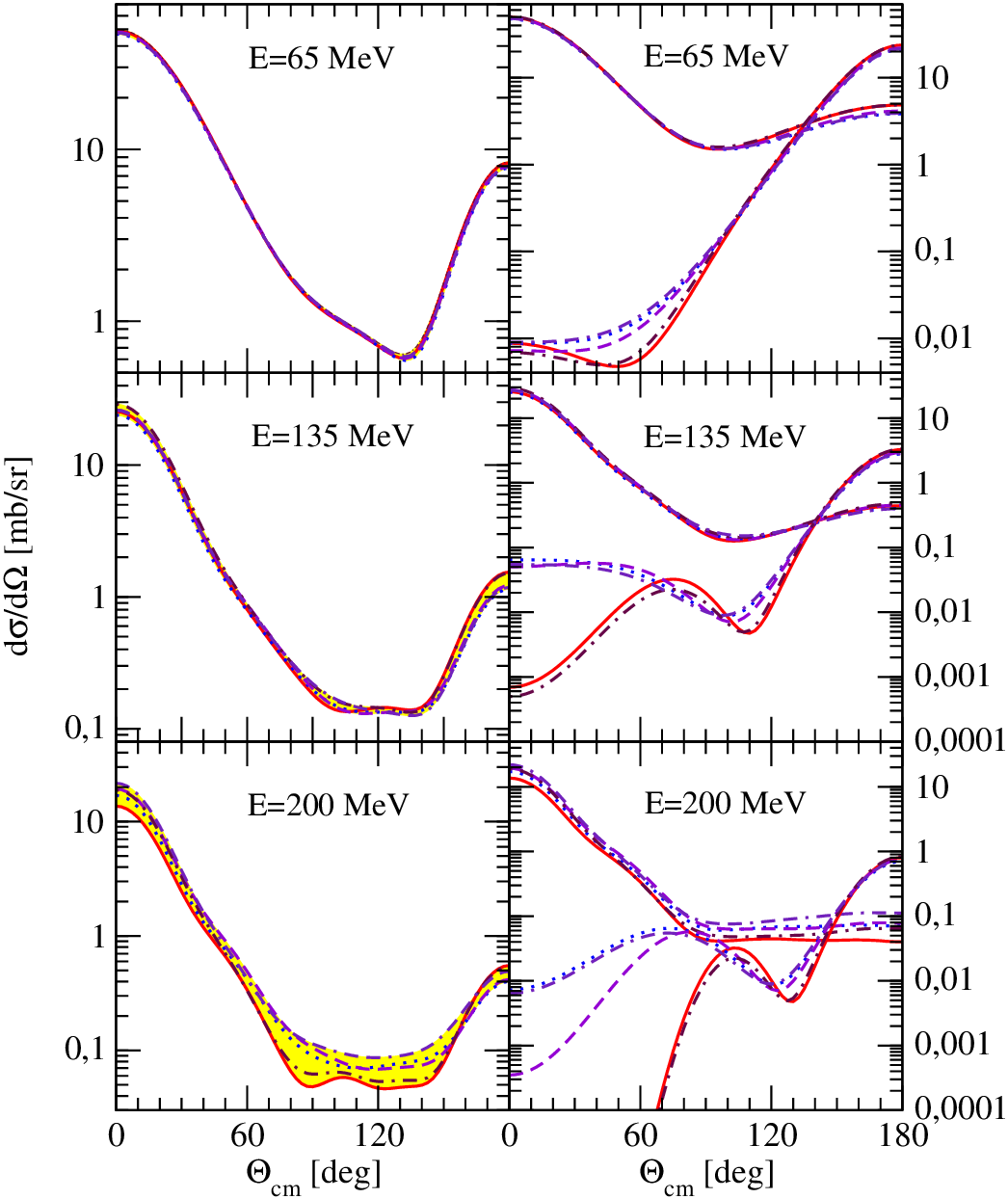}
\caption{
(color online)
The same as in Fig.\ref{fig11} but for the Bochum N$^3$LO NN potentials with 
different cut-off values of Table \ref{table1}. 
See Fig.\ref{fig11} for the description of lines and 
bands.
}
\label{fig12}
\end{figure}

\begin{figure}
\includegraphics[scale=0.7]{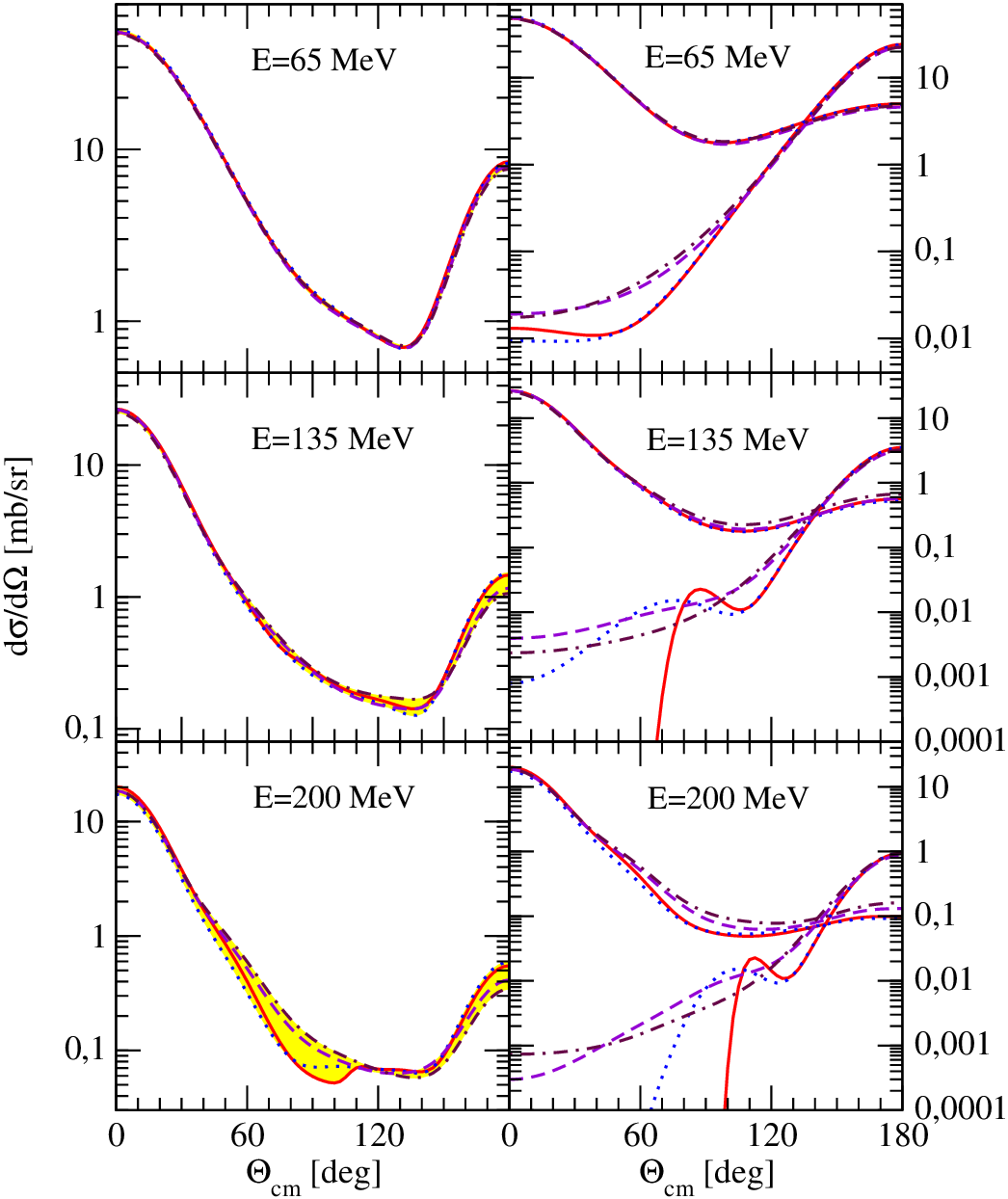}
\caption{
(color online) 
The  nd elastic scattering angular distributions at $65$~MeV, $135$~MeV 
and $200$~MeV 
of incoming neutron lab. energy calculated with  
 the Idaho N$^3$LO NN potentials 
 for different cut-off values of Table \ref{table2}. 
In the left part the nd elastic scattering cross sections are shown. In 
 the right part ``cross sections'' resulting only from the $PT$ term  
in the elastic scattering transition amplitude are shown (lines peaked at 
forward angles)  together with ``cross sections'' based on 
 exchange-term $PG_0^{-1}$ only (lines peaked at backward angles). 
Different lines correspond to different cut-off parameters 
from Table \ref{table2}: 
414 - solid (red), 
450 - dotted (blue), 
500 - dashed (violet), 
600 - dashed-dotted (maroon).  
In the left part light-shaded (yellow) band shows scatter of predictions 
for different cut-off values.  
}
\label{fig13}
\end{figure}

In Figs. \ref{fig11}, \ref{fig12}, and \ref{fig13} in the left column we 
show a band of predictions obtained 
 for nd elastic  scattering cross section at 
those three energies and in the right column ``cross sections'' calculated 
 with the $ \phi'  PT  \phi $ term  (lines peaked at 
forward angles) and with the exchange term $\phi'  PG_0^{-1}  \phi $   
(lines peaked at backward angles) separately. 
At $E_N^{lab}=65$~MeV where $q_0=1.18$~fm$^{-1}$ the ``exchange term'' 
cross sections start to deviate below $\Theta_{cm} \approx 80^o$.
In this angular 
region they are more than one order of magnitude smaller 
 than ``$PT$-term'' 
cross sections. As a consequence a narrow band of predictions for the 
cross section is obtained for all values of the cut-off parameter in the case 
of the Bochum and Idaho N$^3$LO potentials. At $135$~MeV, where 
$q_0=1.7$~fm$^{-1}$, the different cut-off values lead to ``exchange-term'' 
cross sections of quite different angular dependence. It falls down 
drastically with decreasing c.m. angle for smaller cut-off values. 
That behavior is seen even more clearly at $200$~MeV, where 
$q_0=2.07$~fm$^{-1}$. Since with increasing energy at larger c.m. angles 
the exchange-term gets more important compared to the $PT$ term, the 
resulting band of cross section predictions becomes broader, especially 
at $200$~MeV. At $135$~MeV, in spite of the fact that the band is 
rather narrow, one sees also influence of the deuteron wave function 
on the cross section, which leads to different angular behavior of the 
N$^3$LO Bochum and Idaho predictions, especially in the region of the cross 
section minimum. 
 
In Fig.\ref{fig11} we show the predictions of the Bochum N$^2$LO potentials. 
Since N$^2$LO deuteron wave functions do not differ significantly and behave 
similarly to the deuteron wave functions of standard NN forces 
(see Fig.\ref{fig8}) also their influence on the elastic scattering cross 
sections is negligible. The growing band width of N$^2$LO predictions 
reflects the decreasing quality of experimental NN phase-shifts 
description by N$^2$LO forces with increasing energy. Taking N$^3$LO forces 
reduces that width significantly (see Figs. \ref{fig12} and \ref{fig13}).

One could argue that such behavior restricts the application 
 of $\chi$PT forces to 
a rather narrow range of 3N system energies below $\approx 100$~MeV of
the incoming nucleon lab. energy. If that was true, it would make  
the application of the chiral approach impossible in 
 the most interesting region of energies, 
where the consistency between 2N- and 3N-forces plays 
the most important role. 
   One possible way to resolve this problem would be to restrict 
 the range of cut-off to larger values. 
 The other is connected to the omission of 3NF's in above 3N continuum
 calculations. 
 Since neglection 
of a 3NF in nuclear Hamiltonian is an artifact and one should use both 
NN and 3NF's simultaneously, one could argue that when both are applied 
the dependence  on the cut-off parameter would be diminished. 
 To check if that is the case one must perform fully converged
 calculations with 2N and 3N chiral forces included. 
 At higher 
energies much more partial waves are required. 
 The large number of terms contributing to the chiral N$^3$LO
 3NF and huge computer resources needed to calculate their matrix elements
 in higher partial waves preclude presently 
fully converged nd calculations at higher energies. 
 Therefore an approach in which 
partial wave decomposition is avoided and instead vector  Jacobi momenta 
are used is desirable. In the following section we review the
present state of such three-dimensional approach to 3N Faddeev
equations.

\section{Three-dimensional formulations}
\label{3dim}

We restrict ourselves to the momentum space integral equation
formulations.

As already stated, below the pion production threshold the momentum space
Faddeev equations for three-nucleon (3N) scattering can be solved 
with high accuracy 
for essentially all modern two- and three-nucleon forces. In these calculations
angular momentum eigenstates for the two- and three-body systems are used. This
partial wave decomposition (PWD) replaces continuous angular variables 
by a finite set
of orbital angular momentum quantum numbers and allows 
 one to reduce the number 
of continuous variables to only two.

For low projectile energies the procedure of employing 
 orbital angular momentum 
components is well justified. Physics arguments are related to the centrifugal 
barrier and the short range of the nuclear force. Besides, in a numerical 
realization, a fast convergence of the observables with respect to the number
of partial wave states is easily achieved.

However, when considering 3N scattering at higher energies, 
a much larger number of partial wave states is required before results 
become fully convergent. This has obvious consequences for the computational
resources required in the calculations. A considerable effort is required, 
especially when 3N forces with numerous spin-momentum and isospin 
structures are implemented. Preparation of their matrix elements
in the partial wave representation, either analytically or
numerically, is highly nontrivial when the set of partial waves grows rapidly.
Thus it appears natural to avoid PWD and work directly with vector variables. 
We describe below the remarkable progress made in this field.

The main problem in 3N scattering is related to a treatment of the singularity 
in the free 3N propagator above the break-up threshold. The way to deal with 
this difficulty was introduced in \cite{3Nprop1}, where subtraction techniques 
were employed to integrate the logarithmic singularities. For quite 
 a long time 
this approach seemed to be the only possibility and just minor deviations 
from the scheme of \cite{3Nprop1} were introduced. For example, in \cite{12n} 
the logarithmic singularities were integrated quasi-analytically using
splines. Only very recently, a new form of the integral kernel 
for 3N scattering was found, first in the context of the partial 
wave decomposed Faddeev equation
\cite{3Nprop2} and later for its three dimensional (3D) realization in
the 
case of
three bosons \cite{25n}.
In this new approach the treatment of the 3N Faddeev equation becomes 
essentially 
as simple as the treatment of the two-body Lippmann-Schwinger equation.

Without that new insight, it was most natural in the past to proceed also
for the 3D treatments of the 3N Faddeev equation in the following way.
First, the calculation of the deuteron wave function and the two-nucleon (2N) 
scattering t-matrices was required to provide an input to this equation.
Then, typically the Faddeev equation for the 3N bound state was solved. 
In this case the 3N energy is negative, so one encountered no problems 
with the 3N 
propagator. The other ingredients of the formalism could be, 
however, carefully tested.
In particular, the permutation operator and the t-matrices for
negative 
2N energies
had to be well mastered in the bound state calculations.
Finally, the full problem of nucleon-deuteron (Nd) scattering above 
the break-up threshold 
without or with a 3N force could be studied. In this case not only the number 
of variables was bigger due to the presence of the external momentum,
but 
also the 
singularity of the free 3N propagator had to be taken into account.

If one is interested not only in pure $Nd$ scattering
but also in reactions involving 3N scattering states,
like in the description of 
inelastic electron scattering 
 on $^3$H, $ e+ ^3{\rm H} \rightarrow e' + n + d$, 
then it is of great importance 
that the ingredients of the 3D formalism: the deuteron 
 and $^3$H wave functions 
as well as the neutron-deuteron scattering state are prepared consistently.

\subsection{Nucleon-nucleon scattering in three dimensions}
\label{nnsc}

Nucleon-nucleon scattering was treated without PWD already more than 
twenty years ago. 
In \cite{01n1,01n2} the time-dependent Schr\"odinger equation was
solved 
eventually 
for a one-boson exchange potential. It is worth mentioning that 
in the latter paper the general form of the potential 
between two spin-$1/2$ particles was used to simplify the calculations. 
 
Later in \cite{01n4} quasielastic electron 
scattering was investigated and the final-state interaction was taken
into account by evaluating the two-body t-matrix 
directly in 3D for the Malfliet-Tjon (MT III) local spin independent
force \cite{MTIII}. More systematically the angular and
momentum dependence of the t-matrix was studied in the same 
3D approach on as well as off the energy shell in \cite{03n}, 
both for positive and negative 2N energies. In this very informative
paper the behaviour of the t-matrix
in the vicinity of bound-state pole and resonance poles in the second
energy sheet were also investigated for 
 different Malfliet-Tjon-type potentials.

Another alternative to the usual PWD
technique was outlined in \cite{02n}. There 
the two-body Lippmann-Schwinger equation
was written in a numerically solvable form using helicity
theory and taking advantage of the symmetries of the NN interaction.
The numerical examples were based on the Bonn OBEPR potential \cite{obepr}.
The helicity formalism was also used in \cite{06n} (with slightly modified 
final equations) for two quite different NN
potentials, the Bonn~B \cite{bonnb} and 
the Argonne V18 \cite{AV18}. 
The same helicity approach was subsequently used 
by S. Bayegan {\em et al.} \cite{27n}
in 3D calculations of NN bound and scattering states with a chiral 
N3LO potential \cite{epel_nn_n3lo}.
In all these works an excellent agreement with the results
based on standard PWD was reported.

Inclusion of the Coulomb interaction on top of 
a local spin-dependent short-range interaction
in two-body scattering was carried out in \cite{22n}.
The calculations are not performed for the NN system
but their implications are important for all results,
where the screening and renormalization approach is used 
to treat the Coulomb interaction. 

Parallel to the above mentioned nonrelativistic studies,
3D formulations of the scattering equations were studied 
also for the relativistic equations. 
In \cite{11n} this was outlined in the case of pion-nucleon and NN scattering 
treated via the Bethe-Salpeter equation. In \cite{13n} 
a numerical method, based on the Pad\'e summation, was introduced to
solve the covariant spectator equation without partial wave decomposition, and
applied to the NN system.

Last but not least we would like to mention calculations 
of the NN t-matrix, which employ directly
momentum vectors and use spin-momentum operators
multiplied by scalar functions of the
momentum vectors. This approach stems from the fact 
that a general NN force being invariant under time-reversal,
parity, and Galileo transformations can depend only
on six linearly independent spin-momentum operators.
The representation of the NN potential using spin-momentum
operators leads to a system of six coupled equations of scalar
functions (depending on momentum vectors) for the NN
t-matrix, once the spin-momentum operators are analytically
calculated by performing suitable trace operations. This treatment,
formulated in \cite{29n}, can be considered as a natural extension 
for two spin-$1/2$ particles
of the calculations described in \cite{03n},
In \cite{29n} numerical examples for the Bonn~B \cite{bonnb} 
and chiral N2LO potentials \cite{epel_nn_n3lo,n2lo2,n2lo3} 
were presented. Later in \cite{33n} the same approach 
(with a modified choice of the basis spin-momentum operators)
was applied to the Argonne V18 potential \cite{AV18}.
Further variations of this method and inclusion of the Coulomb force
can be found in \cite{31n,35n}. Finally, the application of this 
operator based approach to the
deuteron electro-disintegration process was discussed in \cite{34n}.

\subsection{Few-nucleon bound states in three dimensions}
\label{fewnbs}

We start with the deuteron representations formulated 
without any resort to PWD.

In \cite{07n} the helicity representation developed previously 
for NN scattering \cite{06n} 
was applied to the 2N bound state
and the deuteron eigenvalue equation in the helicity basis
was solved with the Bonn~B potential \cite{bonnb}.
In the same paper 
the deuteron wave function in the so-called (momentum space)
operator form was also derived. In this representation the whole information
about the deuteron is given by two scalar functions,
$\phi_1 (p)$ and $\phi_2 (p)$, which are closely related
to the standard $S$ and $D$ components of the deuteron. 
The direct set of two coupled equations for $\phi_1 (p)$ 
and $\phi_2 (p)$ was derived only later 
in \cite{28n}.
That derivation included simple trace operations, 
which helped eliminate spin degrees of freedom and led 
to analytically given sets of scalar functions depending 
on momentum vectors only.
Numerical examples for the Bonn~B \cite{bonnb} 
and chiral N2LO potentials \cite{epel_nn_n3lo,n2lo2,n2lo3}
were published in \cite{29n}.
Corresponding three dimensional calculations of 2N binding energies 
with chiral N3LO potentials \cite{epel_nn_n3lo}
performed in the helicity formalism were reported in \cite{27n}.

As already mentioned, the work on NN scattering has very often a preparatory
character and further application of the t-matrices are usually planned. 
This is true also in the case of the 3D calculations.
Results of \cite{03n} were later used in \cite{04n}
three-body bound-state calculations without PWD
with Malfliet-Tjon-type NN potentials, neglecting 
spin and isospin degrees of freedom.
In the subsequent paper \cite{08n} the scheme from \cite{04n} was extended
to include scalar two-meson exchange three-body forces.

The Teheran group published several papers dealing with 3D 
solutions of the 3N and even four-nucleon (4N) bound states
\cite{p19n,19n,16n,16.5n,14n,32n,38n,37n}. They started with 
a formulation, which neglected the spin-isospin
degrees of freedom \cite{p19n} and introduced step by step 
improved dynamical ingredients to their framework, performing calculations 
with more realistic NN potentials (like the Bonn~B one in \cite{19n,16n})
and including additionally a 3N force (for example the
Tucson-Melbourne 
3N potential
in \cite{14n}). The 3D t-matrices - an input 
to the systems of coupled equations - were obtained 
with the helicity representation of \cite{06n}.

Finally, we list publications dealing with the 3D treatment 
of the 3N bound state which relies on the general form 
of the 2N t-matrix and the operator form of the 3N bound state
introduced 
in \cite{10n}.
The latter consists of eight operators built from
scalar products of relative momentum and spin vectors, which are applied to 
a pure 3N spin $1/2$ state. Each of the operators is multiplied by 
a scalar function of the relative momentum vectors.
In \cite{28n} one Faddeev equation for identical bosons was replaced 
by a finite set of coupled equations for scalar functions which depend 
only on three variables. The inclusion of a 3N force into this 3D Faddeev 
framework was also discussed.
Further elements of this formalism, for example the construction of the full 
wave function from the Faddeev amplitude, and first numerical results 
for chiral 2N and 3N 
N$^2$LO nuclear forces were provided in \cite{28.5n}.

\subsection{Nucleon-deuteron scattering in three dimensions}
\label{ndsc}

Here we come to the essential, from the point of view of the present
paper, 
3D calculations
of $Nd$ scattering. 

They were initiated in \cite{05n}, soon after the t-matrix \cite{03n}
and the three-body bound  state 
\cite{04n} were successfully treated without employing of PWD. 
Although in \cite{05n} the authors focused on 
scattering below the three-body breakup threshold and avoided the singularity 
of the free three-body propagator, they had to tackle the pole in 
the two-body t-matrix
at the energy corresponding to the two-body bound state. Spin and 
isospin degrees
of freedom were neglected in that pioneering paper.
This framework was naturally extended to higher energies in \cite{12n},
where the Faddeev equation was solved for the three-boson case
using the Pad\'e method. 
The elastic differential cross section, semiexclusive $d(N,N')$ 
 cross sections, 
and the total cross sections for the elastic and breakup processes at energies 
up to about 1 GeV were calculated.

At such high projectile energies 
special relativity is expected to become
relevant. The key advantage of a formulation of the Faddeev
equations in terms of vector variables lies in the fact 
that using them, the relativistic three-body problem can be formulated 
more easily than  in PWD scheme. 
To achieve this goal, in \cite{15n} authors worked within the
framework of Poincar\'e invariant quantum mechanics,
where the dynamical equations have the same number
of variables as the corresponding nonrelativistic equations.
The equations used to describe the relativistic few-body problem have even
the same operator form as the nonrelativistic ones but their ingredients are
different. This was discussed in detail in \cite{15n} and the authors 
restricted themselves to the leading-order
term of the Faddeev multiple scattering series within the
framework of Poincar\'e invariant quantum mechanics, since it 
already contained many relativistic ingredients. In that paper
kinematical and dynamical relativistic effects on selected observables 
were systematically studied.
As later found in \cite{20n,17n,25n,26n}, the truncation 
of the multiple scattering series to the first term was not always justified. 
This insight became possible, when the full solution to the relativistic 
Faddeev equation for three bosons had been obtained. Only then
convergence of the multiple scattering series could be investigated 
as a function of the projectile energy in different scattering
observables 
and configurations. 
Based on a Malfliet-Tjon interaction, which generated the same 
nonrelativistic and relativistic 
observables in the 2N system, observables for elastic and breakup 
scattering were calculated and compared
to non-relativistic ones. That gave first insight 
to ``relativistic effects'' in $Nd$
scattering. 
It is also worth noticing that the cross sections 
calculated in \cite{23n} at 500 MeV are in quite good agreement with 
experimental data, in spite of the simple nature of the model interaction.
One may hope that in the future a similar framework will be built
based on 
realistic 
NN forces.

A completely different path was chosen in \cite{09n}. There only the
first 
term of the multiple
scattering series generated by the Faddeev equations was considered
but 
authors used realistic 
NN forces, the Bonn~B \cite{bonnb} and the Argonne V18 \cite{AV18} potentials.
Including spin degrees of freedom, it was possible to calculate some 
polarization observables
for the $d(p, n)pp$ reaction and compare them with the results of PWD 
calculations.
The NN t-matrix elements were calculated in the helicity representation 
defined in \cite{06n}. Based on the information available from calculations 
using PWD, this framework has to be extended towards full solution of 
the Faddeev equation.
Also 3N forces have to be included in this scheme.

We are not aware of a regular paper from the Teheran group about their 
treatment of
$Nd$ scattering in 3D, although photodisintegration of $^3$H based on 
realistic NN 
interactions with full inclusion of final 
 state interactions among the outgoing 
nucleons reported in \cite{36n}, requires the same technical achievements 
as $Nd$ scattering.

From our point of view, a very promising way to proceed towards 
(at least) nonrelativistic 3D
treatment of $Nd$ scattering was formulated in \cite{30n}. 
The recently developed ``operator'' formalism for
two- and three-nucleon bound states in three
dimensions \cite{28n,29n,28.5n} was extended to the realm of 
nucleon-deuteron scattering.
The aim was here to formulate the momentum space Faddeev
equations in such a fashion that the final equations would be reduced
to spin-independent scalar functions in the same
spirit as already worked out for the 2N and 3N bound
states. The numerical realization of this framework is in progress.

\section{Summary and outlook}
\label{summary}

Recent derivation of the  N$^3$LO 3NF opened the possibility 
 to test consistent 
two- and three-nucleon chiral forces  
at that order in few-nucleon applications. Since with the growing order of 
the chiral expansion the number of terms contributing 
 to the chiral 3NF increases 
significantly, a prerequisite to such applications was a development 
of fast and efficient method of automated partial wave decomposition. 
Using that method we applied for the first time 
 chiral N$^3$LO 3NF (still without 
 the short-range $2\pi$-contact term amd 1/m relativistic corrections)
to low energy nd elastic scattering 
 and breakup. 
 It turns out that N$^3$LO 3NF cannot explain the low-energy $A_y$-puzzle. 
That result indicates possible drawbacks of low-energy $^3P_j$ NN 
phase-shifts or/and on the lack of some spin-isospin-momenta  structures
 in N$^3$LO 3NF. The standard chiral perturbation theory 
 formulation based on pions 
and nucleons as the only explicit degrees of freedom  still misses  
 at N$^3$LO some physics associated with intermediate $\Delta$ excitations, 
 which can 
be to some extent accounted for only at N$^4$LO order \cite{krebs1,krebs2}. 
 Since $\Delta$ provides an important mechanism leading to a new form of 3NF,
 in the next step one should  apply  that recently 
 derived N$^4$LO 3NF \cite{krebs1,krebs2} in calculations of nd reactions.

Higher-energy nd reactions, in which clear evidence for large 3NF effects 
 was found, call for a 
three-dimensional treatment of the 3N Faddeev equations. The rather 
 simple form of 2- 
and 3-nucleon chiral forces, when expressed in 
 terms of Jacobi momentum vectors, 
encourages approaches which altogether avoid partial wave 
decomposition. Studies of the cut-off dependence of  
N$^3$LO NN chiral interaction in higher-energy nd elastic scattering 
 revealed preference for larger cut-off values. The use of 
 lower cut-offs would 
preclude applications of N$^3$LO chiral dynamics in that interesting region 
of energies.

\section*{Acknowledgments}
Thw authors would like to thank Prof. E. Epelbaum for very
constructive and valuable discussions. 
This work was supported by the Polish National Science Center 
 under Grant No. DEC-2011/01/B/ST2/00578. 
 It was also partially supported 
 by the Japan Society for Promotion of Science (JSPS ID No. S-12028),
 and by the European Community-Research Infrastructure
Integrating Activity
``Exciting Physics Of Strong Interactions'' (acronym WP4 EPOS)
under the Seventh Framework Programme of EU.
 The numerical
calculations have been performed on the
 supercomputer cluster of the JSC, J\"ulich, Germany and Ohio
 Supercomputer Centre, USA (Project PAS0680).
 We acknowledge support by the Foundation for Polish 
 Science - MPD program, co-financed 
by the European Union within the Regional Development Fund.

\vfill
\newpage

\vfill
\newpage

\vfill
\newpage

\vfill
\newpage

\vfill
\newpage

\vfill
\newpage

\vfill
\newpage

\newpage

\vfill
\newpage

\vfill
\newpage

\vfill
\newpage

\vfill
\newpage

\vfill.
\newpage

\vfill
\newpage

\begin{table}
\begin{tabular}{|c|c|c|c|}
\hline
 cut-off & ($\Lambda$, $\tilde{\Lambda})  $   & $c_D$ & $c_E$ \\
         &  [MeV]                             &      &        \\
\hline
1 & (450,500) & 10.78 & -0.172 \\
2 & (600,500) & 12.00  & 1.254  \\
3 & (550,600) & 11.67 & 2.120   \\
4 & (450,700) & 7.21  & -0.748 \\
5 & (600,700) & 14.07 & 1.704 \\
\hline
\end{tabular}
\caption{The values of $c_D$ and $c_E$ LECs for  the Bochum N$^3$LO 
   potentials 
with different cut-off parameters shown in the second column. In the N$^3$LO 
3NF relativistic $1/m$ corrections and 
 the $2\pi$-exchange-contact term were omitted.} 
\label{table1}
\end{table}

\begin{table}
\begin{tabular}{|c|c|c|c|c|}
\hline
cut-off & 1 & 2 & 3 & 4 \\
\hline
$\Lambda$ [MeV] & 414 & 450 & 500 & 600  \\
\hline
\end{tabular}
\caption{The values of the cut-off parameter $\Lambda$ for the different
  N$^3$LO  Idaho potentials.}
\label{table2}
\end{table}

\end{document}